\documentclass[reprint,amsmath,amssymb,aps]{revtex4-2}
\usepackage{stackengine,tikz,
    float,graphicx,hyperref,bm,dcolumn,tabularx}

\newcommand{\xdownarrow}[1]{%
  {\left\downarrow\vbox to #1{}\right.\kern-\nulldelimiterspace}
}

\newcommand\stackequal[2]{%
  \mathrel{\stackunder[2pt]{\stackon[4pt]{=}{$\scriptscriptstyle#1$}}{%
  $\scriptscriptstyle#2$}}}

\begin{document}
\title{Evidences of the Generalizations of BKT Transition in Quantum Clock Model}
\author{Bingnan Zhang}%
 \email{bingnan.zhang@rutgers.edu}
\affiliation{Department of Physics, Rutgers New Brunswick, NJ 08854, USA}
\begin{abstract}
We calculate the ground state energy density $\epsilon(g)$ for the one dimensional N-state quantum clock model up to order 18, where $g$ is the coupling and $N=3,4,5,...,10,20$.   Using methods based on Pad\'e approximation, we extract the singular structure of $\epsilon''(g)$ or $\epsilon(g)$. They correspond to the specific heat and free energy of the classical 2D clock model\cite{Suzu}. We find that, for $N=3,4$, there is a single critical point at $g_c=1$.The heat capacity exponent of the corresponding 2D classical model is $\alpha=0.34\pm0.01$ for $N=3$, and $\alpha=-0.01\pm 0.01$ for $N=4$. For $N>4$, There are two exponential singularities related by $g_{c1}=1/g_{c2}$, and $\epsilon(g)$ behaves as $Ae^{-\frac{c}{|g_c-g|^{\sigma}}}+analytic\ terms$ near $g_c$. The exponent $\sigma$ gradually  grows from $0.2$ to $0.5$ as N increases from 5 to 9, and it stabilizes at 0.5 when $N>9$. These phase transitions should be generalizations of Kosterlitz-Thouless transition, which has $\sigma=0.5$. The physical pictures of these phase  transitions are still unclear.
\end{abstract}
\maketitle

\section{Introduction}\par
The classical 2D N-state clock model is a generalization of the Ising model. When $N=2$, it {\it is} the Ising model, and it becomes the XY model when $N\rightarrow\infty$. 
It is widely believed that for $N\leq 4$ there is a second order phase transition as we dial up the temperature from zero, while there are two BKT-like transitions for $N>4$\cite{Elit}\cite{Einh}\cite{Hame}\cite{Nien}\cite{Froh}. However, the quantitative behavior of the free energy for $N>4$ is still controversial . In an early paper \cite{Elit}, the singular part of the free energy was argued to behave like $e^{-\frac{c}{|T_c-T|^{\sigma}}}$ near $T_c$, where $\sigma\approx0.22$ for $N=5$, $\sigma\approx0.5$ for $N>5$. More recent simulations indicate that $\sigma=0.5$ for  $N=5$ \cite{Bori}. There are also simulations that claim to show that the transition in $N=5$ model is not BKT-like \cite{Baek}. In this paper, we will try to shed some light on these questions by studying the  1D quantum clock model, whose ground state energy energy maps to the free energy of the corresponding 2D classical model \cite{Suzu}. 
\par
The paper is organized as the follows: section \ref{Model} specifies the model. Section \ref{LCE} discusses the linked cluster expansion, which is the method used in calculating the series. The codes used in this section can be downloaded at \url{https://github.com/beyondoubt3/clock-model-perturbation}. Section \ref{Fit} introduces Pad\'e approximation and its improvements. Section \ref{Results} includes the series and fitting results and section \ref{Summary} is a summary. Readers who are familiar with the linked cluster expansion, Pad\'e approximations and inhomogeneous differential approximations can go directly to section \ref{Results}.
 
\section{The model}\label{Model}\par
\begin{figure}[H]
\begin{center}
\begin{tikzpicture}
\draw[thick,->](0,0)--(9,0)node[pos=0.1,below]{i-1}node[pos=0.5,below]{i}
node[pos=0.9,below]{i+1}
node[pos=0.1,above]{\includegraphics[scale=0.1]{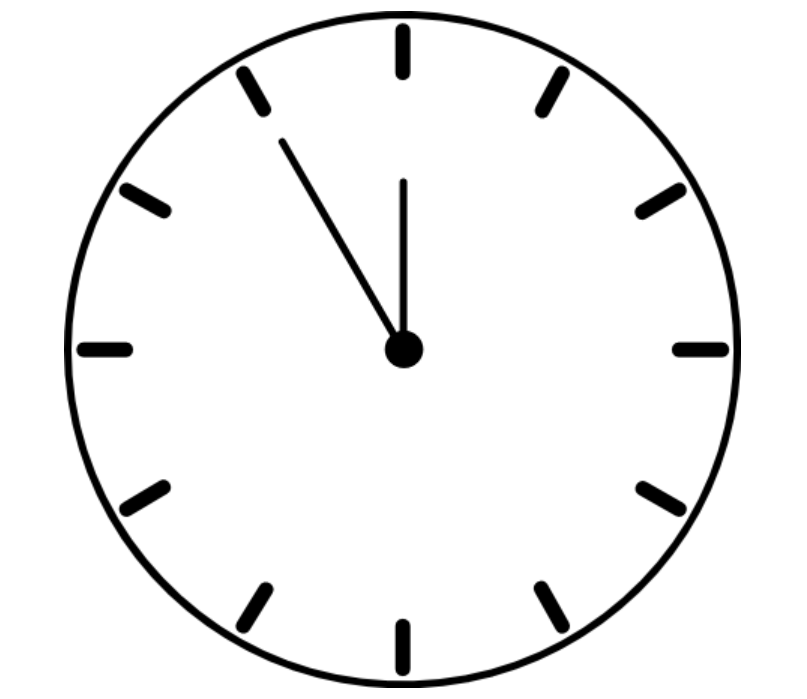}}
node[pos=0.5,above]{\includegraphics[scale=0.1]{image.png}}
node[pos=0.9,above]{\includegraphics[scale=0.1]{image.png}}
;
\end{tikzpicture}
\end{center}
\caption{The clock model.}
\label{fig:clock}
\end{figure}

The Hamiltonian is 
\begin{equation}
H=-g\sum_i(V_i+V_i^{\dagger})-\sum_i(U_iU_{i+1}^{\dagger}+U_{i+1}U_i^{\dagger})
\label{eq:H}
\end{equation}
where $i$ runs over all sites on the one dimensional lattice, and we use periodic boundary conditions.
\begin{equation}
V_i=
\begin{pmatrix}
0&1&0&...&0\\
0&0&1&...&0\\
\vdots&\vdots&\vdots&\ &\vdots \\
0&0&0&...&1\\
1&0&0&...&0
\end{pmatrix}
\ \ 
U_i=
\begin{pmatrix}
1&0&0&...&0\\
0&e^{\frac{2\pi i}{N}}&0&...&0\\
0&0&e^{\frac{4\pi i}{N}}&...&0\\
\vdots&\vdots&\vdots&\ &\vdots \\
0&0&0&...&e^{\frac{2\pi(N-1)i}{N}}
\end{pmatrix}
\end{equation}
are generators of the finite Heisenberg group, and they satisfy
\begin{equation}
V_iU_i=\omega U_iV_i,\ \omega=e^{i\frac{2\pi}{N}}
\end{equation}
 Working in the basis  where $U_i$ operators are diagonal, the effect of $U_i$ is to read the needle's position on the $i$th clock, and $V_i$ dials up  the needle by one unit (figure \ref{fig:clock}).  

In the limit $g\rightarrow 0$, every clock's needle is locked at the same position, and there is a $N-fold$ degeneracy. The perturbation won't mix any two of them when the order of perturbation theory is small compared to the lattice size, so we can arbitrarily choose one as our starting point. Once the starting point is chosen, we can relabel the states by the difference in position between two nearest neighbor clocks. To be more precise, define $u_i,\ v_i $ such that 
\begin{equation}
u_i=U_iU_{i+1}^{\dagger},\ V_i=v_{i-1}^{\dagger}v_i
\end{equation}   
where $u_i,\ v_i$ have the same matrix representations as $U_i,\ V_i$. Figure \ref{dualitypic} is an example of state relabeling. The first line labels the state by eigenvalues of $U_i$ while the second line labels the state by eigenvalues of $u_i$:
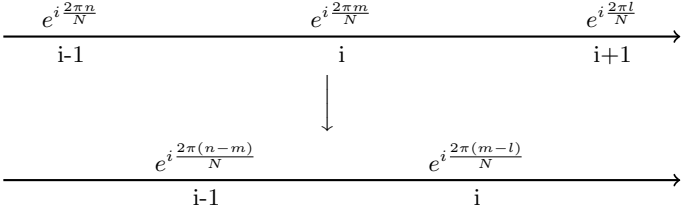
\begin{figure}[H]
\begin{center}
\begin{tikzpicture}
\draw[thick,->](0,0)--(9,0)
node[pos=0.1,below]{i-1}
node[pos=0.5,below]{i}
node[pos=0.9,below]{i+1}
node[pos=0.1,above]{$e^{i\frac{2\pi n}{N}}$}
node[pos=0.5,above]{$e^{i\frac{2\pi m}{N}}$}
node[pos=0.9,above]{$e^{i\frac{2\pi l}{N}}$};
\end{tikzpicture}
\\
$\bigg\downarrow$\\
\begin{tikzpicture}
\draw[thick,->](0,0)--(9,0)
node[pos=0.3,below]{i-1}
node[pos=0.7,below]{i}
node[pos=0.3,above]{$e^{i\frac{2\pi (n-m)}{N}}$}
node[pos=0.7,above]{$e^{i\frac{2\pi (m-l)}{N}}$};
\end{tikzpicture}
\end{center}
\caption{State relabeling.}
\label{dualitypic}
\end{figure}
Now we do a second relabeling. Since 
\begin{equation}
v_iu_i=\omega u_iv_i,\ \ u_i^{\dagger}v_i=\omega v_iu_i^{\dagger}
\end{equation}
We can relabel
\begin{equation}
u_i\rightarrow v_i,\ \ v_i\rightarrow u_i^{\dagger}
\end{equation}
After these two steps, the Hamiltonian becomes
\begin{equation}
\begin{aligned}
H(g)&=-g\sum_i(u_iu_{i+1}^{\dagger}+h.c.)-\sum_i(v_i+h.c.)\\&
=g\left(
-\sum_i(u_iu_{i+1}^{\dagger}+h.c.)-\frac{1}{g}\sum_i(v_i+h.c.)
\right)
\end{aligned}
\end{equation}
Thus the ground state energy density at large $g$ and small $g$ are related by 
\begin{equation}
\epsilon(g)=g\epsilon(\frac{1}{g})
\label{eq:duality}
\end{equation}

\section{Calculating the series}\label{LCE}
\par The key method used here is linked cluster expansion\cite{Oitm}. As an analog of Feynman diagram expansion in Field theory, linked cluster expansion states that the perturbation series of ground state energy, or any extensive quantities for an Hamiltonian lattice system, only receives contributions from connected clusters of lattice sites. 
\begin{equation}
E(G)=\sum_{G'\in G}b(G'/G)e(G')
\label{eq:E}
\end{equation}
where $G$ is the cluster we are interested in,  $G'$ runs over all sub-clusters in $G$, and $b(G'/G)$ is the embedding number that tells us how many ways to embed cluster $G'$ within $G$. $e(G')$ is the 'reduced energy'. It is different from $E(G')$ because $E(G')$ receive contributions from all of $G'$'s sub-clusters while $e(G')$ only counts $G'$ itself. 
\par The good thing about linked cluster expansion is that for any finite cluster $G'$, the series for $e(G')$ starts from an order proportional to the cluster size. For example, in our model (\ref{eq:H}), the clusters are just chains with certain lengths. If $G'$ has length $k$, then $e(G')$ starts from order $2k$. This means that if we want to calculate $E(G)$ to order $2k$, we only have to pay attention to the sub-clusters of $G$ with size smaller than $k+1$.
\par In many cases, the embedding number $b$ is very difficult to evaluate. However, for the 1D lattice considered in this paper, the embedding number is extremely simple. Let us denote by $G_n$ a chain with length $n$.  There are $m-n+1$ different  ways to embed a chain of length  $n$ within a chain of length $m$ ($m>n$ of course), so the embedding number of $G_n$ inside $G_m$ is
$$b(G_n/G_m)=m-n+1$$
so equation (\ref{eq:E}) becomes
\begin{equation}
E(G_m)=me(G_1)+(m-1)e(G_2)+(m-2)e(G_3)+...+e(G_{m})
\end{equation}
If we are only interested in series below order $2k$ ($k<m$), We can truncate the above series 
\begin{equation}
\begin{aligned}
E(G_m)\stackrel{to\ order\ 2k}{=}&me(G_1)+(m-1)e(G_2)+(m-2)e(G_3)+...\\&+(m-k+1)e(G_{k})
\end{aligned}
\end{equation} 
In this paper, we are interested in the ground state energy density in the thermodynamic limit, so 
\begin{equation}
\begin{aligned}
\frac{E(G_m)}{m}\stackequal{to\ order\ 2k}{m\rightarrow\infty}&e(G_1)+e(G_2)+e(G_3)+...+e(G_{k})
\\=&E(G_{k})-E(G_{k-1})
\end{aligned}
\end{equation} 
\par Now the only thing left is to do perturbations on chains with length $k$ and $k-1$ to calculate $E(G_k)$ and $E(G_{k-1})$. Here we use the standard Rayleigh-Schr\"odinger perturbation theory. We expand the ground state energy
\begin{equation}
E^0=E^0_0+E^0_1+E^0_2+...
\end{equation}
and the ground state wavefunction
\begin{equation}
|\Psi^0\rangle=|\Psi^0_0\rangle+\sum_{j>0}c_{j,1}|\Psi^j_0\rangle+\sum_{j>0}c_{j,2}|\Psi^j_0\rangle+...
\end{equation}
$E^0_j$ and $c_{j,r}$ can be calculated with  iteration\cite{Oitm}. 
\begin{equation}
\begin{aligned}
&E_r^0=\sum_j H'_{0,j}c_{j,r-1}\\
&c_{j,r}=\frac{1}{E_0^j-E_0^0}(-\sum_We H'_{j,i}c_{i,r-1}+\sum_{s=0}^{r-1}E_{r-s}^0c_{j,s}),j\neq 0
\end{aligned}
\label{eq:iteration}
\end{equation}
where $H'_{j,i}=\langle \Psi^j_0|H'|\Psi^i_0\rangle$ and $H'$ is the first order part of the Hamiltonian. 
\par The input data is the set of states $|\Psi^j_0\rangle$. One doesn't have to generate all states in the Hilbert space, because many of them won't be used. If we want to calculate the series to order $2k$, only those $|\Psi^j_0\rangle$ that satisfy $\langle \Psi^j_0|(H')^a|\Psi^0_0\rangle\neq 0$ for a certain integer $0<a\leq k$ will be used. Once the states are generated, the calculation of $E_0^j$ is fast and straightforward. Calculating $H'_{j,i}$ is, however, time-consuming. In fact, most of the time in this algorithm is spent on calculating $H'_{j,i}$. 
The reason is that in order to calculate $H'_{j,i}$, a simple algorithm will run through the states twice, roughly speaking. We believe that using some formulas in restricted partition theory, there is a clever way to sort the states such that one only have to run through the states once. If this is true, significantly more orders can be obtained. The codes used in this section can be downloaded online at \url{https://github.com/beyondoubt3/clock-model-perturbation}.

\section{Fitting methods}\label{Fit}
\subsection{Pad\'e and DLog Pad\'e}
\par
Instead of fitting an unknown function by its series expansion, which is a polynomial, the Pad\'e approximation\cite{Bake} fits the function by the quotient of two polynomials.  For example, suppose
\begin{equation}
f(g)=f_0+f_1g+f_2g^2+...+f_sg^s+O(g^{s+1}),\ g<1
\end{equation}
 we fit $f(g)$ by the form
\begin{equation}
f_{[n/m]}(g)=\frac{\sum_{i=0}^n p_ig^i}{1+\sum_{j=1}^m q_jg^j}
\end{equation}
where $n+m+1=s+1$. One solves for $p_i,q_j$ by requiring that $f=f_{[n/m]}$ below order $s+1$. Pad\'e fitting often yield good results because the large $g$ behavior is controlled. It is very good at capturing poles and zeros within the convergence region, and it will mimic a branch cut by accumulating poles and zeros along that line. However, sometimes fake poles will appear to render the approximation inaccurate, so only those poles and zeros which are stable when we change the orders $m,n$ of the approximant should be trusted.. In principle, $n$ and $m$ can be any positive integer. However, what happens most is that the fitting works best when $n$ and $m$ are close. In many cases, one just sets $n=m$.
\par If we have the expansions of $f(g)$ at two points, we let $f_{[n/m]}$ agree with those two series simultaneously, and this is called two point Pad\'e. Of course, one of the points can be infinity.
\par DLog Pad\'e is used in extracting critical exponents. Suppose $f(g)$ goes like $(g_c-g)^{-v}$ near $g_c$, then $\frac{dlog(|f|)}{dg}=\frac{f'}{f}$ goes like $\frac{-v}{g-g_c}$ near $g_c$. Since Pad\'e approximant is particularly good at fitting poles of order one, we can estimate $-v$ by applying Pad\'e approximation to $\frac{f'}{f}$ and then calculate its residue at $g_c$.

\subsection{Inhomogeneous differential approximation\\(IDA)}\label{IDA}
\par Sometimes DLog Pad\'e gives a bad estimate of critical exponents. This is caused by background terms, and IDA\cite{Fish} takes that into account. We write $f(g)=A(g)(1-\frac{g}{g_c})^{-v}+B(g)$ near $g_c$, where $A(g)$ and $B(g)$ are both analytic. $f(g)$ satisfies
\begin{equation}
u(g)+p(g)f(g)-q(g)f'(g)=0
\label{eq:IDA}
\end{equation}
where
\begin{equation}
\begin{aligned}
u(g)=&(g_c-g)A(g)B'(g)-[vA(g)+(g_c-g)A'(g)]B(g)\\
p(g)=&vA(g)+(g_c-g)A'(g)\\
q(g)=&(g_c-g)A(g)
\end{aligned}
\end{equation}
\par In practical use, we choose $u,p,q$ to be polynomials , and solve equation(\ref{eq:IDA}) order by order. $g_c$ can be estimated as the stable zero point of $q$, and $v$ can be estimated as $\frac{-p'(g_c)}{q(g_c)}$. When we have both small $g$ and large $g$ series for $f(g)$, we fix the order of $u,p,q$ by letting the three terms in equation(\ref{eq:IDA}) start at the same order, both in the small $g$ and large $g$ limits.  We've found that without these restrictions the results are unreliable.

\section{Results}\label{Results}\par
The ground state energy density series in the small $g$ limit are listed in appendix \ref{C}. The series in the large $g$ limit can be obtained by the duality relation (\ref{eq:duality}).

\subsection{N=3,4}
\par One can map the 1D quantum model to a 2D classical model, and $\epsilon(g)$ becomes the free energy, $\epsilon''(g)$ becomes specific heat. It is believed that for $N=3,4$ there is a second order phase transition\cite{Elit}. Table \ref{dlog1},\ref{dlog2} are the results of applying single side DLog Pad\'e fitting to $\epsilon''(g)$ for $N=3,4$:\\

\begin{table}[H]
\begin{ruledtabular}
\begin{tabular}{lcr}
Polynomial type&Pole&Residue\\
\colrule
$[5/6]$&1.00644&-0.465541
\\
$[4/5]$&1.01735&-0.510407
\\
$[3/4]$&1.01595&-0.507258
\\
$[2/3]$&1.0369&-0.556233
\\
\end{tabular}
\end{ruledtabular}
\caption{N=3 fitting $\epsilon^{'''}/\epsilon^{''}$}
\label{dlog1}
\end{table}

\begin{table}[H]
\begin{ruledtabular}
\begin{tabular}{lcr}
Polynomial type&Pole&Residue\\
\colrule
$[6/6]$&1.00951&-0.281564
\\
$[5/5]$&1.01481&-0.307984
\\
$[4/4]$&1.02066 &-0.329759
\\
$[3/3]$&1.04277&-0.387972
\\
\end{tabular}
\end{ruledtabular}
\caption{N=4 fitting $\epsilon^{'''}/\epsilon^{''}$}
\label{dlog2}
\end{table}
In both cases the pole position stabilizes  
at 1, but the residue is still changing. This is the effect of background terms. In order to improve the estimation of the residue, we use IDA (section \ref{IDA}). The results are listed below (table \ref{ida1},\ref{ida2}):
\begin{table}[H]
\begin{ruledtabular}
\begin{tabular}{lcr}
Polynomial type&Pole&Residue\\
\colrule
u2 p5 q6&1.00027&-0.339898
\\
u4 p4 q5&1.00025&-0.338126\\
u1 p4 q5&1.00098&-0.353952
\end{tabular}
\end{ruledtabular}
\caption{N=3 IDA fitting $\epsilon^{'''}/\epsilon^{''}$}
\label{ida1}
\end{table}

\begin{table}[H]
\begin{ruledtabular}
\begin{tabular}{lcr}
Polynomial type&Pole&Residue\\
\colrule
u2 p5 q6&0.999829&$0.00716088$
\\
u5 p4 q4&0.999509&$0.019595$\\
u4 p5 q4&0.999637&$0.0148813$
\end{tabular}
\end{ruledtabular}
\caption{N=4 IDA fitting $\epsilon^{'''}/\epsilon^{''}$}
\label{ida2}
\end{table}
'u5 p8 q9' means we set $$u=\sum_{i=0}^{5}u_ig^i, p=\sum_{i=0}^{8}p_ig^i, q=1+\sum_{i=1}^{9}q_ig^i$$.
\par A summary of the above results: both $N=3,N=4$ models have a single critical point at $g_c=1$. The specific heat exponent $\alpha$ for the corresponding 2D classical model is estimated as $0.34\pm0.01$ for $N=3$, and $-0.01\pm0.01$ for $N=4$.  As we know, the 3-state clock model is equivalent to the 3-state Potts model\cite{Wu}, and the 4-state clock model is equivalent to two copies of the Ising model at the critical point\cite{Bett}. Both of them are exactly solvable:$\alpha=1/3$ for $N=3$ and $\alpha=0$ for $N=4$\cite{Wu}. Comparing to these exact results, our estimates are fairly reasonable. We notice that in an early paper \cite{Enti}, the series for the 3-state model was calculated to order 31, and $\alpha$ was estimated with DLog Pad\'e method. Their results never stabilize, which is consistent with what we found at first. The IDA method is thus a significant improvement. 
\par One can further refine the result for $N=4$ by using the two point IDA, using both small g series and large g series. However, the two point method doesn't work very well for the $N=3$ model.
\subsection{N$>$4}\label{N>4}
\par
For $N>4$ we expect to see two essential singularities related by $g_{c_1}=1/g_{c_2}$,  and the singular part of free energy is $e^{-\frac{c}{|g_c-g|^{\sigma}}}$\cite{Elit}. When $N\rightarrow \infty$ the model maps to the classical XY model in 2D, $\sigma=0.5$. In order to find the critical point, we take a derivative. $\epsilon'(g)$ approaches zero like $-\frac{A}{(g_c-g)^{\sigma}}e^{-\frac{c}{(g_c-g)^{\sigma}}}$ near $g_c$. Fitting $\epsilon'$ with Pad\'e polynomials, we indeed found a stable zero point. We have tried a number of different methods for unveiling an exponential singularity masked by an analytic background, and in the end this simple method worked the best.  In appendix \ref{A} we apply this method to an explicitly known function and show that it works well.

\begin{figure}[H]
\centering
\includegraphics[scale=0.42]{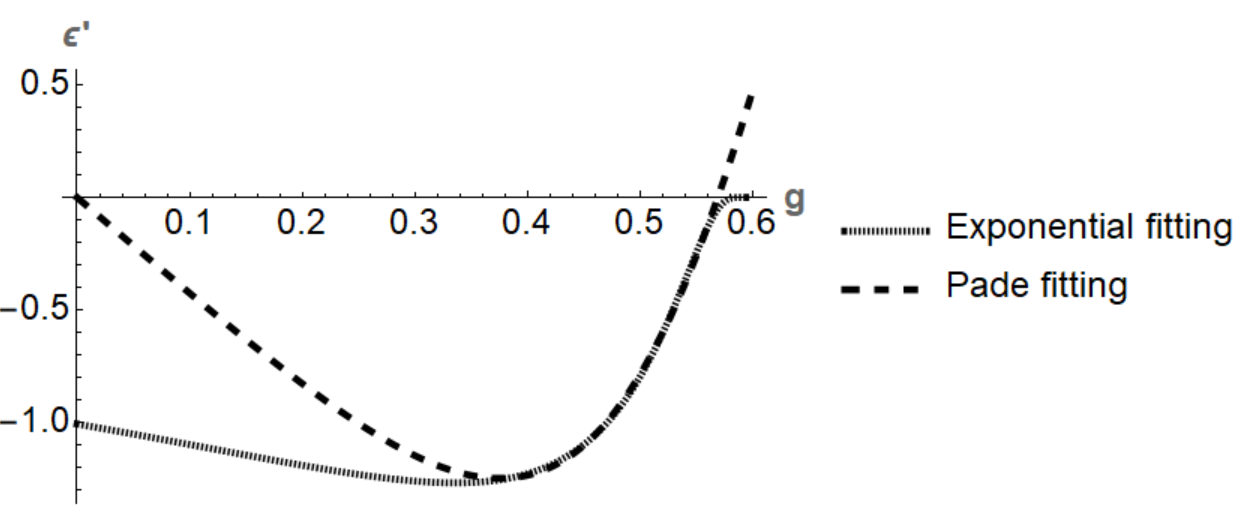}
\caption{$\epsilon'$ for $N=9$}
\label{epN9}
\end{figure}

The upper curve in figure \ref{epN9} is the Pad\'e fitting result for $N=9$. The qualitative results for other $N>4$ cases are similar.
The region with positive slope on the left hand side of $g_c$ should be dominated by the singular structure. We fit this part of the curve with $-\frac{A}{(g_c-g)^{\sigma}}e^{-\frac{c}{(g_c-g)^{\sigma}}}$ to extract $g_c$ and $\sigma$. The lower curve in figure \ref{epN9} is the fitting result for $N=9$, and the numerical results are summarized in the table below (table \ref{main}): 
\begin{table}[H]
\begin{ruledtabular}
\begin{tabular}{lcr}
N&$\sigma$&$g_c$\\
\colrule
5&$0.21\pm 0.01$&$1.1\pm 0.2$\\
6&$0.29\pm 0.05$&$0.96\pm0.07$\\
7&$0.31\pm0.05$&$0.83\pm 0.05$\\
8&$0.40\pm0.05$&$0.68\pm0.05$\\
9&$0.5\pm 0.1$&$0.57\pm 0.05$\\
10&$0.5\pm0.1$&$0.48\pm0.05$\\
20&$0.5\pm 0.1$&$0.136\pm0.015$
\end{tabular}
\end{ruledtabular}
\caption{Estimations of $\sigma$ and $g_c$.}
\label{main}
\end{table}

\par The error is obtained by changing the fitting region slightly or changing the order of the Pad\'e polynomials. There is a clear trend that $\sigma$ increases from 0.2 to 0.5 as we dial up $N$. The results also show that $g_c<1$ for $N>5$. According to the duality relation (\ref{eq:duality}), there should be another singularity $1/g_c$ with the same singular behavior. For $N=5$, our results are not precise enough to tell whether $g_c<1$ or not. However, Pad\'e fitting to $\epsilon(g)$ and $\epsilon'(g)$ both give two stable poles in the $g>0$ region (appendix \ref{B}), although they don't obey the duality relation (\ref{eq:duality}). More sophisticated methods are required to find the precise location of $g_c$ for $N = 5$.
\subsection{Evidence that $N\leq4$ and $N>4$ are different}
\par One may feel uncomfortable with our fitting method, because we manually separate the cases $N\leq4$ and $N>4$. In this subsection we present some evidence that they really belong to two classes. \par
\begin{figure}[H]
\centering
\includegraphics[scale=0.52]{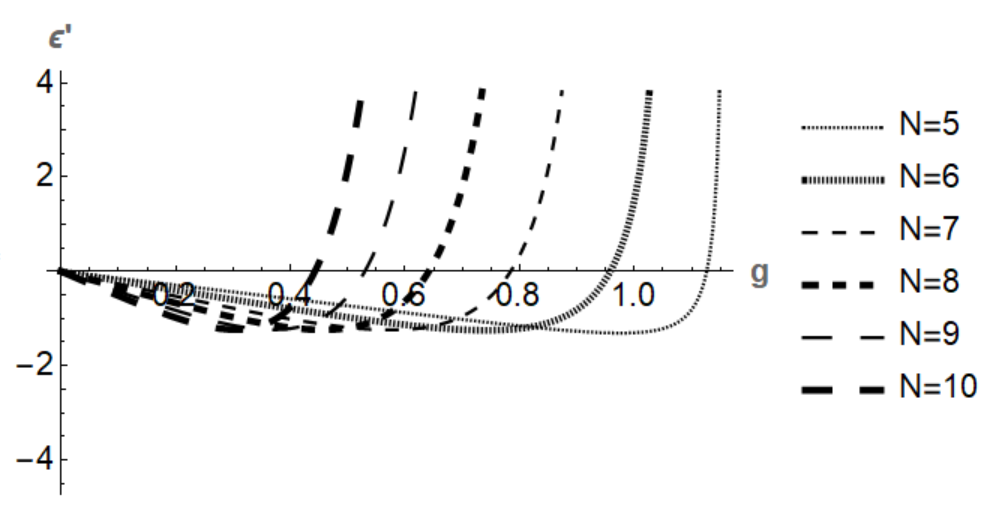}
\caption{$\epsilon'$ for $N>4$}
\label{fig:N5678910ep}
\end{figure}
\begin{figure}[H]
\centering
\includegraphics[scale=0.52]{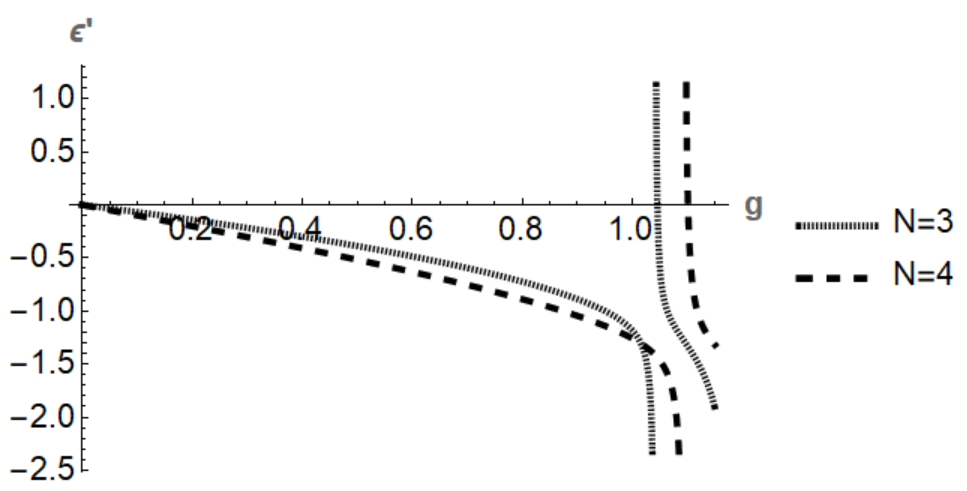}
\caption{$\epsilon'$ for $N=3,4$}
\label{fig:N34ep}
\end{figure}

The first piece of evidence is the  behavior of the fit to  $\epsilon'$ (figure \ref{fig:N5678910ep},\ref{fig:N34ep}). When $N>4$, it goes to 0 at $g_c$ to mimic exponential suppression. When $N\leq4$, it diverges because poles begin to accumulate in the $g>g_c$ region, telling us that there is a branch cut.

\par
\begin{figure}[H]
\centering
\includegraphics[scale=0.6]{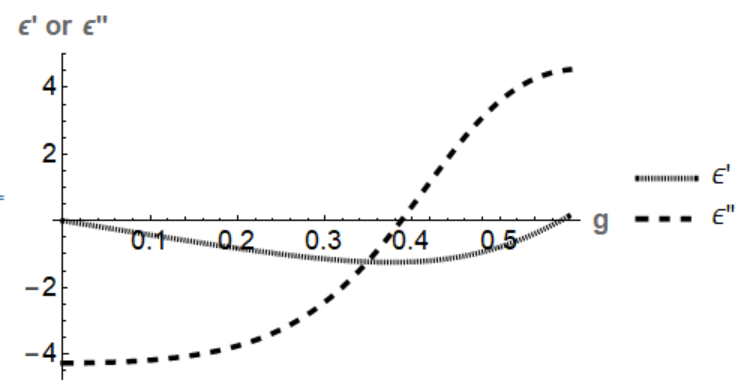}
\caption{The first zero point of $\epsilon''$ is smaller than that of $\epsilon'$ for N=9.
 The qualitative behaviors are the same for $N>5$.}
\label{fig:N9epepp}
\end{figure}

 As a second test, we applied Pad\'e fitting to $\epsilon''$, and found a stable zero point for $N>5$ that is smaller than the zero point of $\epsilon'$ (figure \ref{fig:N9epepp}). This is consistent with an exponential singularity $e^{-c/(g_c-g)^{\sigma}}$, because as we take more derivatives to the exponential form, smaller and smaller zeros will appear on the left hand side of $g_c$. This will not happen for a singularity that goes like $(g_c-g)^{-\sigma}$. Instead, poles and zeros will accumulate on the right hand side of $g_c$, and this is exactly what happened when we fit $\epsilon''$ for $N=3,4$.  It is also these additional zeros that make the Dlog Pad\'e or inhomogeneous differential approximation to $\epsilon''/\epsilon'$ inaccurate, because it seems that Pad\'e approximants only work for $g$ smaller than the first zero point. For $N=5$, however, there is no clear signature that $\epsilon''$ has a smaller zero point than $\epsilon'$.

\section{Summary}\label{Summary}
\par We calculated the ground state energy of the 1-dimensional N-state quantum clock model up to order 18, and extracted its singular structure near the critical point, for values of $N$ up to $20$. It was found that, for $N=3,4$, there is a single critical point at $g_c=1$, and the exponent $\alpha$ for the corresponding 2D classical model is $0.34\pm0.01$ for $N=3$ while that for $N=4$ is $-0.01\pm0.01$. For $N>4$, There are two exponential singularities related by $g_{c1}=1/g_{c2}$, and the ground state energy behaves as $Ae^{-\frac{c}{|g_c-g|^{\sigma}}}+analytic\ terms$ near $g_c$. The exponent $\sigma$ gradually  grows from $0.2$ to $0.5$ as N increases from 5 to 9. 
These findings show that there exist a class of generalizations of KT transition, and more insights can be obtained by studying models with $N=5,6,7,8$. Better methods of extracting   exponential singularities are also needed to find the precise transition point for $N=5$, and to get a better estimation of the exponents. We also need physical pictures for these novel phase transitions.

\section{Acknowledgments}
\par The author would like to thank Professor Tom Banks for countless discussions throughout the research, and Professor Rajiv Singh, Professor Peter Young who taught the author about the linked cluster expansion. This research was supported by DOE under grant number 
DOE-SC0010008.

\appendix
\section{Fitting the exact series of an exponential singularity}\label{A}
\par Here we repeat the calculation in section \ref{N>4}, with the series of $\epsilon$ replaced by the series of $e^{-\frac{1}{(g_c-g)^{\sigma}}}$. The results are listed below:
\begin{table}[H]
\begin{ruledtabular}
\begin{tabular}{lcdr}
\textrm{$g_c$}&\textrm{$g_{c fit}$}&\textrm{$\sigma$}&\textrm{$\sigma_{fit}$}\\
\colrule
1&1.0&0.5&0.53\\
0.8&0.80&0.4&0.38\\
0.6&0.59&0.3&0.16\\
\end{tabular}
\end{ruledtabular}
\end{table}
If we replace $\epsilon$ with $sin(g)+e^{-\frac{1}{(g_c-g)^{\sigma}}}$, where $sin(g)$ represents an analytic background term, the results are:
\begin{table}[H]
\begin{ruledtabular}
\begin{tabular}{lcdr}
\textrm{$g_c$}&\textrm{$g_{c fit}$}&\textrm{$\sigma$}&\textrm{$\sigma_{fit}$}\\
\colrule
1&0.94&0.5&0.43\\

0.9&0.89&0.4&0.45\\

0.8&0.79&0.3&0.18\\
\end{tabular}
\end{ruledtabular}
\end{table}

We see that the fitting results for $g_c$ are always precise. $|g_c-g_{cfit}|$ is controlled below $0.06$. The estimations of $\sigma$ when $\sigma\geq 0.4$ are also reasonable. $|\sigma-\sigma_{fit}|$ is always smaller than $0.07$. When $\sigma=0.3$, the error can be as large as $0.14$. But $\sigma_{fit}$ is still smaller than $0.5$, so it's qualitatively correct.
\section{Stable real poles of $\epsilon(g)$ and $\epsilon'(g)$ when $N=5$}\label{B}
\begin{table}[H]
\begin{ruledtabular}
\begin{tabular}{cc}
\textrm{Polynomial type}&
\textrm{Real poles for $\epsilon(g)$}\\
\colrule
$[9/8]$&$0.788095, 1.24853, 5.67655$\\
$[8/7]$&$1.26307,5.14201$\\
$[7/6]$&$1.21815, 7.88399$\\
$[5/4]$&$1.28659, 6.82486$\\
\end{tabular}
\end{ruledtabular}
\end{table}

\begin{table}[H]
\begin{ruledtabular}
\begin{tabular}{cc}
\textrm{Polynomial type}&
\textrm{Real poles for $\epsilon'(g)$}\\
\colrule
$[7/7]$&$1.15988, 2.88669$\\
$[6/6]$&$1.0942, 4.79821$\\
$[5/5]$&$1.33875, 3.8271$\\
$[4/4]$&$1.16288, 4.26156$\\
\end{tabular}
\end{ruledtabular}
\end{table}

\section{The ground state energy density series}\label{C}
\begin{table}[H]
\begin{ruledtabular}
\begin{tabular}{cc}
\textrm{$N$}&
\textrm{$\epsilon(g)$}\\
\colrule
3&
$
\begin{aligned}
&-2-0.33333333333333326g^2
-0.05555555555555558g^3\\&- 
 0.015432098765432216g^4-
 0.010802469135802482g^5\\&-
0.005596517299192159g^6- 0.003382783789733934g^7\\&-0.0024418478191840687g^8- 0.0017083823549243165g^9\\&- 
0.0012505362271225252g^{10}- 0.0009640588459594025g^{11}\\&- 0.0007532607066100003g^{12}-0.0006002324824848607g^{13}\\& - 0.0004884680877595293g^{14} - 
 0.00040306911349643267g^{15}\\& - 0.0003367619172180523g^{16} - 
 0.00028468719829159695g^{17}\\& - 0.00024306955918608876g^{18}+O(g^{19})
\end{aligned}
$
\end{tabular}
\end{ruledtabular}
\end{table}

\begin{table}[H]
\begin{ruledtabular}
\begin{tabular}{cc}
\textrm{$N$}&
\textrm{$\epsilon(g)$}\\
\colrule
4&
$
\begin{aligned}
&-2-0.5g^2-0.03125g^4-0.0078125g^6-
0.0030517578125g^8\\&- 0.001495361328125g^{10}-
0.0008411407470703125g^{12}\\&- 0.0005192756652832031g^{14}-0.00034280307590961456g^{16}\\&+O(g^{18})
\end{aligned}
$\\
\\
5&
$
\begin{aligned}
&-2-0.7236067977499783g^2 -0.01082711823295851g^4\\&
-0.0050000000000000044g^5+ 0.012834208342438508g^6\\&
-0.004476748935007664g^7+ 0.0071691845190831605g^8\\&
-0.0014582789036703438g^9+ 0.00363340974879511g^{10}\\&+
0.00042817768624983307g^{11}+ 0.0009825430279915533g^{12}\\&+
0.0009009222496709941g^{13}+ 0.00027667263772020514g^{14}\\&+
0.0009072543398714415g^{15} -0.00006787109149526856g^{16}\\&
+0.00048470861631427255g^{17} + 0.00004878354324601734g^{18}\\&+O(g^{19})
\end{aligned}
$\\
\\
6&
$
\begin{aligned}
&-2-g^2+0.016666666666666607g^4+ 0.08043055555555534g^6\\&+
0.05265773483846181g^8+ 0.03615288217802676g^{10}\\&+
0.005796349140238799g^{12} -0.005592712308918113g^{14}\\&
-0.04275193393460419g^{16}+O(g^{18})
\end{aligned}
$\\
\\
7&
$
\begin{aligned}
&-2-1.3279852776056806g^2+ 0.09797290679170478g^4\\&+
0.35397193583896414g^6 -0.000637755102040817g^7\\& +0.3085285764405121g^8
-0.0019474457234286945g^9\\& +0.2608242736968025g^{10}
-0.0008936678382355442g^{11}\\& -0.27307294524474823g^{12}
+0.006680757899727427g^{13}\\& -0.5706053033136786g^{14} +0.014034959344389308g^{15}\\&
-3.1048948786406925g^{16}
+0.030755900001374098g^{17}\\&
 - 0.6577812813212631g^{18}+O(g^{19})
\end{aligned}
$\\
\\
8&
$
\begin{aligned}
&-2-1.7071067811865461g^2+ 0.28545145311140274g^4\\&+
1.2452271184234416g^6+ 1.4021698692988345g^8\\&+
1.481704469767327g^{10} -6.297416226458864g^{12}\\&
-13.136066464760518g^{14} -119.4409371149577g^{16}+O(g^{18})
\end{aligned}
$\\
\\
9&
$
\begin{aligned}
&-2-2.1371580426032555g^2+ 0.6607776898982003g^4\\&+
3.7994617012802543g^6+ 5.479144293201969g^8\\&
-0.00009645061728395061g^{9}+ 7.090071901275373g^{10}\\&
-0.0008103680532492554g^{11} -84.32489227044425g^{12}\\&
-0.0008946489050342989g^{13} -200.28503141812962g^{14}\\& 0.02226349726079538g^{15}
-3201.6764082094887g^{16}\\&
+0.08036690433538807g^{17} + 6989.784090313808g^{18}+O(g^{19})
\end{aligned}
$\\
\\
10&
$
\begin{aligned}
&-2-2.618033988749893g^2+ 1.3441206615821706g^4\\&+
10.379000385996093g^6+ 18.939873150481205g^8\\&+
29.18326653883338g^{10} -829.172485374158g^{12}\\&
-2284.117236667422g^{14} -63633.110395201365g^{16}+O(g^{18})
\end{aligned}
$\\
\\
20&$
\begin{aligned}
&
-2-10.215864547265355g^2+ 103.49335226791732g^4\\&+
8906.588873287546g^6+ 103245.9724322232g^8\\&- 
 212594.28296551108g^{10} - 2820319368.8904266g^{12}\\& - 
 23222204677.125977g^{14} - 41202014867887.25g^{16}+O(g^{18})
\end{aligned}
$

\end{tabular}
\end{ruledtabular}
\end{table}


\begin{thebibliography}{99}
\bibitem{Elit}E.Rabinovici, R. B. Pearson, and J. Shigemitsu. Phase structure of discrete Abelian spin and gauge systems. Physical Review D 19,3698(1979).
\bibitem{Einh}M.B.Einhorn, R.Savit, and E.Rabinovici. A physical picture for the phase transitions in zn symmetric models.Nuclear Physics B 170,16(1980).
\bibitem{Hame} C.J.Hamer and J.B.Kogut. Weak-coupling series and the critical indices of Z p spin systems in two dimensions. Physical Review B 22,3378 (1980).
\bibitem{Nien}B.Nienhuis,Critical behavior of two-dimensional spin models and charge asymmetry in the Coulomb gas. Journal of Statistical Physics 34,731(1984).
\bibitem{Froh}J.Frohlich, and T.Spencer. The Kosterlitz-Thouless transition in two-dimensional abelian spin systems and the Coulomb gas. Communications in Mathematical Physics 81,527(1981).
\bibitem{Oitm}J.Oitmaa,H.Chris, and W.Zheng. Series expansion methods for strongly interacting lattice models. Cambridge University Press, 2006. 
\bibitem{Bake}G.A.Baker and J.L.Gammel. The Padé approximant in theoretical physics. Academic Press, 1970.
\bibitem{Fish}M.E.Fisher and H.A.Yang. Inhomogeneous differential approximants for power series. Journal of Physics A: Mathematical and General 12,1677(1979).
\bibitem{Enti}I. G.Enting,Series analysis for the three-state Potts model. Journal of Physics A: Mathematical and General 13,L133(1980).
\bibitem{Bori}O.Borisenko et al. Numerical study of the phase transitions in the two-dimensional Z (5) vector model. Physical Review E 83,041120(2011).
\bibitem{Baek}S.K.Baek and M.Petter. Non-Kosterlitz-Thouless transitions for the q-state clock models. Physical Review E 82,031102(2010).
\bibitem{Suzu}M.Suzuki,Relationship between d-dimensional quantal spin systems and (d+ 1)-dimensional ising systems: Equivalence, critical exponents and systematic approximants of the partition function and spin correlations. Progress of theoretical physics 56,1454(1976).
\bibitem{Wu}F.Y.Wu, The potts model. Reviews of modern physics 54,235(1982).
\bibitem{Bett}D.D.Betts, The exact Solution of some Lattice Statistics Models with four States per Site. Canadian Journal of Physics 42,1564(1964).
\end{thebibliography}
\end{document}